\documentclass[twocolumn]{aastex63}

\begin{document}

\title{Dynamical structure of small bulges reveals their early formation in $\Lambda$CDM paradigm}

\correspondingauthor{Luca Costantin}
\email{l.costantin@cab.inta-csic.es}

\author[0000-0001-6820-0015]{Luca Costantin}
\affiliation{Centro de Astrobiolog\'ia (CSIC-INTA), Ctra de Ajalvir km 4, Torrej\'on de Ardoz, E-28850 Madrid, Spain}
\affiliation{INAF - Astronomical Observatory of Brera, via Brera 28, I-20121
Milano, Italy}

\author{Jairo M\'endez-Abreu}
\affiliation{Instituto de Astrof\'isica de Canarias, Calle Vía L\'actea s/n, E-38200 La Laguna, Tenerife, Spain}
\affiliation{Departamento de Astrof\'isica, Universidad de La Laguna, Calle Astrof\'isico Francisco S\'anchez s/n, E-38205 La Laguna, Tenerife, Spain}

\author{Enrico M.~Corsini}
\affiliation{Dipartimento di Fisica e Astronomia ``G. Galilei'', Universit\`a di Padova, vicolo dell'Osservatorio 3, I-35122, Padova, Italy}
\affiliation{INAF - Osservatorio Astronomico di Padova, vicolo dell'Osservatorio 5, I-35122 Padova, Italy}

\author{Lorenzo Morelli}
\affiliation{Instituto de Astronom\'ia y Ciencias Planetarias, Universidad de Atacama, Copayapu 485, Copiap\'o, Chile}

\author{Adriana de Lorenzo-C\'aceres}
\affiliation{Instituto de Astrof\'isica de Canarias, Calle Vía L\'actea s/n, E-38200 La Laguna, Tenerife, Spain}
\affiliation{Departamento de Astrof\'isica, Universidad de La Laguna, Calle Astrof\'isico Francisco S\'anchez s/n, E-38205 La Laguna, Tenerife, Spain}

\author{Ilaria Pagotto}
\affiliation{Leibniz-Institut f\"ur Astrophysik Potsdam (AIP), An der Sternwarte 16, D-14482 Potsdam, Germany}

\author{Virginia Cuomo}
\affiliation{Dipartimento di Fisica e Astronomia ``G. Galilei'', Universit\`a di Padova, vicolo dell'Osservatorio 3, I-35122, Padova, Italy}

\author{J. Alfonso L. Aguerri}
\affiliation{Instituto de Astrof\'isica de Canarias, Calle Vía L\'actea s/n, E-38200 La Laguna, Tenerife, Spain}
\affiliation{Departamento de Astrof\'isica, Universidad de La Laguna, Calle Astrof\'isico Francisco S\'anchez s/n, E-38205 La Laguna, Tenerife, Spain}

\author{Michela Rubino}
\affiliation{Dipartimento di Fisica e Astronomia ``G. Galilei'', Universit\`a di Padova, vicolo dell'Osservatorio 3, I-35122, Padova, Italy}

\begin{abstract}

The $\Lambda$ cold dark matter ($\Lambda$CDM) paradigm of galaxy formation predicts that dense 
spheroidal stellar structures invariably grow at early cosmic time.
These primordial spheroids evolve toward a virialized dynamical status as they
finally become today's elliptical galaxies and large bulges at the center of disk galaxies.
However, observations reveal that small bulges in spiral galaxies are common in the nearby universe.
The prevailing belief that all small bulges form at later times from internal processes occurring in the disk 
represents a challenge for the $\Lambda$CDM scenario.
Notably, the coevolution of bulges and central supermassive black holes (SMBHs) at early phases of galaxy evolution
is also at stake.
However, observations have so far not provided conclusive evidence against their possible early origin. 
Here, we report new observations of small bulges showing that they follow the mass--velocity dispersion relation 
expected for virialized systems. Contrary to previous claims, 
small bulges bridge the gap between massive ellipticals and globular clusters. This dynamical
picture supports a scenario where systems over seven orders of magnitude in stellar mass form at early cosmic time.
These results alleviate the tension between $\Lambda$CDM simulations and observations at galactic scales. 
We hypothesize that these small bulges are actually the low-mass descendants of compact objects observed 
at high redshift, also known as red nuggets, which are consistently produced in cosmological $\Lambda$CDM simulations. 
Therefore, this also suggests that the established coevolution of SMBHs and 
large bulges naturally extends to spheroids in the low-mass regime.
\end{abstract}


\section{Introduction \label{sec:intro}}

The $\Lambda$CDM model is very successful at reproducing observations 
of the large-scale structures of the universe in a cosmological context.
In this scenario, dark matter halos initially convert their potential energy into kinetic energy
reaching a virialized status, allowing then galaxies to form due to the cooling of baryons \citep{White1978}.
However, the evolution of luminous structures is observed to markedly deviate 
from the assembly history of dark matter halos, challenging
our understanding of the interplay between baryonic structures and the dark matter 
content in the universe \citep{Bullock2017}.

The degree by which galaxies satisfy fundamental scaling relations likely reflects
the underlying dynamical principles of how baryons settle within dark matter potential wells.
The theoretical prediction from the Virial Theorem is that
the photometric and kinematic information of virialized systems are coupled in a
way that they tend to occupy a preferential position in fundamental scaling relations
\citep[e.g., Fundamental Plane;][]{Djorgovski1987}.

In this context, the brightest and most massive early-type galaxies (ETGs) 
were the first systems observed to satisfy the fundamental scaling relations valid for virialized systems.
The evolution of ETGs relies on the so-called two-phase formation scenario \citep{Oser2010}. 
In the first stage, the galaxy forms the bulk of its stellar component experiencing  
a gas-rich dissipational collapse within dark matter halos.
In the second stage, from redshift $z\lesssim2$, dry accretion of satellites and occasional major mergers
drive their evolution.
Regardless of the mechanisms that funnel the gas to the center of the dark halos
(e.g., disk instabilities, accretion from cosmic web, or wet mergers),
the initial phase of galaxy formation leads to a dense stellar core in a rapid formation process, 
known as gas compaction \citep{Dekel2009}.
The build-up of the core is regulated by the growth of the central supermassive black hole (SMBH) that coevolves with it \citep{Silk1998}.
This process, which takes place at high redshift, sets the overall properties of massive ETGs, 
which ultimately translate into the scaling relations observed in the local universe after the system virialization process.

The compelling evidence that large bulges swelling out the center of massive spiral galaxies
satisfy the same fundamental scaling relations of ETGs leads us to infer that they share a similar formation process.
This scenario implies an early formation of the bulge, prior to the disk, 
that would be instead  strongly perturbed if not destroyed in case of a later formation.
Moreover, observations of SMBHs ubiquitously residing in all massive bulges, 
with bigger black holes living in larger bulges \citep{Ferrarese2000}, 
lead to the belief that black hole growth and bulge formation undeniably regulate each other. 
Interestingly, some authors have proposed that SMBHs do not coevolve with dark matter, disks, 
or disk-like structures within galaxies \citep{Kormendy2011, Kormendy2011b}.

In late-type galaxies secular processes connected to disk evolution
are thought to be responsible for building their small and less massive bulges \citep{Kormendy2004}.
These continuous and slow processes are thought to shape the overall physical properties
of the central region of late-type galaxies into a disk-like structure 
\citep[][but see \citealt{Guo2019} for the build-up of non-disky bulges out of internal secular evolution]{Kormendy2008}.
While ETGs virialized out of the density fluctuations of cold dark matter (CDM) in short timescales,
slow processes responsible for building up disk-like systems prevent them from reaching a virialized status.
Thus, the severe implication is that the proposed disky nature of small bulges 
implies that they have to deviate from the fundamental virial relations valid for ETGs and massive bulges \citep{Fisher2016}.
Moreover, the internal redistribution of angular momentum in the disk-like component
and the gradual gas infall provide less black hole (BH) feeding and may drive slower SMBH growth.
Therefore, the supposed late and slow formation of small bulges in late-type galaxies fuels the tension
between SMBH and bulge coevolution \citep{Kormendy2011, Kormendy2011b}.

\section{Sample and observations} \label{sec:sample}

New spectroscopic observations of a sample of 26 extremely late-type spiral galaxies 
allow us to describe the dynamical status of small bulges in the context of $\Lambda$CDM paradigm.

The sample galaxies have been selected from the Sloan Digital Sky Survey (SDSS) Data Release 13 spectroscopic catalog 
\citep{Albareti2017},
which provides a unique framework for statistical galaxy studies. They are a representative subsample ($\sim10\%$)
of a volume limited sample of 247 galaxies down to $M_r < - 18$ mag located in a volume of 100 Mpc radius.
They are defined as late-type according to: 
$(i)$ concentration $C = R_{90}/R_{50} < 2.5$, typical of late-type disk galaxies; 
$(ii)$ Petrosian radius $R_{\rm petro} > 10''$, avoiding extremely small galaxies where spatial resolution problems could arise.
They also present: $(iii)$ inclination $i < 65^{\circ}$ and $(iv)$ central average surface 
brightness in $r$-band $\mu_r$($r < 2''$) $<$ 20 mag arcsec$^{-2}$, 
allowing for a good photometric definition of the half-light radius of the bulge and limiting the effects of dust lanes.

The spectroscopic observations were carried out in two separate observing runs 
on 2018 February 13-16 (A36TAC\_11/2017B; PI: L.~Costantin) and 2018 May 21-24 (CAT18A\_61/2018A; PI: J.~M\'endez-Abreu).
We used the DOLORES spectrograph at the Telescopio Nazionale Galileo in La Palma (Spain) mounting the V510 grism,
covering the wavelength range \mbox{$4875-5325$ \AA} ~using the $1''$ wide slit.
The spectrograph was equipped with a E2V
4240 camera and a thinned back-illuminated, deep-depleted,
Astro-BB coated charge-coupled device (CCD) with $2048 \times 2048$ pixels.
The instrumental resolution is measured as the mean value of Gaussian FWHMs (Full Width at Half Maximum) 
using a number of unblended arc-lamp lines over the whole spectra range of a wavelength-calibrated spectrum.
We obtain ${\rm FWHM}_{\rm inst} = 0.7$ \AA, corresponding to a
velocity dispersion $\sigma_{\rm inst} = 19$ km s$^{-1}$ at 5100 \AA.
The angular sampling is 0.252 arcsec pixel$^{-1}$, with a reciprocal dispersion
of 0.235 \AA ~pixel$^{-1}$. The median value of the seeing FWHM during
the observing nights was $\sim0''.9$.

\begin{deluxetable}{ccccc}
\tablenum{1}
\tablecaption{Physical parameters of sample bulges.\label{tab:params}}
\tablewidth{0pt}
\tablehead{
\colhead{Galaxy} & \colhead{$r_{\rm e}$} & \colhead{$\sigma_{\rm e}$} & \colhead{$M_{\rm bulge}$} & \colhead{$B/D(r<r_{\rm e})$} \\
\colhead{ } & \colhead{(pc)} & \colhead{(km s$^{-1}$)} & \colhead{(log($M/M_{\odot}$))} & \colhead{ }
}
\decimalcolnumbers
\startdata
NGC~2503          	&   637 $\pm$ 27         &    63 $\pm$ 6            &   9.18 $\pm$ 0.23		     		&	3.6				\\
NGC~2684             	&   225 $\pm$ 7           &    53 $\pm$ 5            &   8.34 $\pm$ 0.22			     	&	1.5				\\
NGC~2742              	&   253 $\pm$ 7           &    53 $\pm$ 3            &   8.21 $\pm$ 0.19			     	&	1.7				\\
NGC~2913              	&   363 $\pm$ 11         &    59 $\pm$ 3            &   8.76 $\pm$ 0.23		     		&	5.7				\\
NGC~3041              	&   354 $\pm$ 24         &    58 $\pm$ 3            &   8.68 $\pm$ 0.22		     		&	2.1				\\
NGC~3055              	&   168 $\pm$ 7           &    50 $\pm$ 4            &   8.20 $\pm$ 0.21		     		&	3.8				\\
NGC~3294              	&   277 $\pm$ 8           &    76 $\pm$ 5            &   8.64 $\pm$ 0.18		     		&	2.8				\\
NGC~3486              	&   144 $\pm$ 4           &    66 $\pm$ 1            &   8.63 $\pm$ 0.18			     	&	4.0				\\
NGC~3810              	&   151 $\pm$ 4           &    65 $\pm$ 1            &   8.55 $\pm$ 0.19			     	&	1.5				\\
NGC~3982              	&   117 $\pm$ 4           &    76 $\pm$ 2            &   8.39 $\pm$ 0.19			     	&	1.3				\\
NGC~4237              	&   123 $\pm$ 4           &    55 $\pm$ 2            &   8.05 $\pm$ 0.19			     	&	1.0				\\
NGC~4305              	&   332 $\pm$ 22         &    42 $\pm$ 1            &   8.68 $\pm$ 0.20		     		&	4.8				\\
NGC~4405              	&   198 $\pm$ 19         &    41 $\pm$ 3            &   8.30 $\pm$ 0.20		     		&	0.7				\\
NGC~4598             	&   431 $\pm$ 43         &    39 $\pm$ 1            &   8.80 $\pm$ 0.20		     		&	1.8				\\
NGC~4620              	&   859 $\pm$ 58         &    38 $\pm$ 1            &   8.66 $\pm$ 0.21		     		&	1.6				\\
NGC~4654              	&   125 $\pm$ 13         &    44 $\pm$ 5            &   8.10 $\pm$ 0.21		     		&	5.0				\\
NGC~4689              	&   664 $\pm$ 44         &    31 $\pm$ 4            &   9.13 $\pm$ 0.19			     	&	1.5				\\
NGC~5480              	&   267 $\pm$ 8           &    57 $\pm$ 5            &   8.61 $\pm$ 0.21			     	&	3.9				\\
NGC~5768              	&   441 $\pm$ 13         &    64 $\pm$ 4            &   8.93 $\pm$ 0.22		     		&	1.8				\\
NGC~5956              	&   414 $\pm$ 27         &    77 $\pm$ 2            &   9.33 $\pm$ 0.21                        	&	6.0				\\
NGC~5957              	&   504 $\pm$ 50         &    77 $\pm$ 4            &   9.11 $\pm$ 0.21			     	&	3.6				\\
NGC~6106              	&   465 $\pm$ 13         &    59 $\pm$ 5            &   7.85 $\pm$ 0.21			     	&	0.4				\\
NGC~7137              	&   130 $\pm$ 4           &    24 $\pm$ 5            &   8.11 $\pm$ 0.20			     	&	3.2				\\
IC~0492                  	&   617 $\pm$ 26         &    86 $\pm$ 6            &   8.25 $\pm$ 0.21		     		&	4.9				\\
IC~1066                  	&   309 $\pm$ 9           &    58 $\pm$ 4            &   9.24 $\pm$ 0.22		     		&	0.5				\\
IC~2293                  	&   213 $\pm$ 9           &    20 $\pm$ 8            &   8.15 $\pm$ 0.23		     		&	45.3				\\
\enddata
\end{deluxetable}

\section{Data analysis \label{sec:analysis}}

We performed standard reduction for both photometric and spectroscopic data, 
as thoroughly described in \citet{Costantin2017}.

For all 26 galaxies, we retrieved $i$-band SDSS DR13 images, in order to properly characterize
the different substructures in the galaxies (i.e., bulge, disk, and bar component).
We derived the structural parameters of the sample galaxies by performing a two-dimensional 
photometric decomposition of their surface brightness distribution using the 
GAlaxy Surface Photometry 2 Dimensional Decomposition
\citep[GASP2D;][]{MendezAbreu2008, MendezAbreu2014}.
In particular, this allows us to properly define the half-light radius of the bulge component, 
which results critical in homogenizing the kinematics measurements.

Following \citet{Taylor2011}, we use the empirical relation between ($g - i$) color, $i$-band luminosity, and stellar mass
\begin{equation}
{\rm log(M_{\star}/}M_{\odot}) = -0.68 + 0.7(g-i) -0.4(\mathcal{M}_i - 4.58)  \, , 
\end{equation}
to derive the stellar mass of the bulge, where $\mathcal{M}_i$ corresponds to its absolute magnitude 
in the rest-frame $i$-band expressed in the AB system. 
The relation provides estimate to a 1$\sigma$ accuracy of $\sim0.1$ dex and was calibrated according
to Galaxy And Mass Assembly (GAMA) sample of galaxies \citep[][]{Driver2009}.

We used the absorption features in the long-slit spectra to measured 
the stellar kinematics within the bulge half-light radius, 
adopting the penalized pixel-fitting method \citep[][]{Cappellari2004} and
the Gas and Absorption Line Fitting algorithm \citep[GANDALF;][]{Sarzi2006}.
The observed spectra were convolved with a linear combination of stellar spectra
from the ELODIE library \citep[FWHM = 0.48 \AA;][]{Prugniel2001}.

The high spectral and spatial resolution kinematics allows us to measure reliable stellar velocity dispersions 
within the bulges half-light radii down to $\sim 20$ km s$^{-1}$, pushing the limit of 
available instruments to a still unexplored regime.
The photometric information from $i$-band SDSS images
allows us to characterize bulges of typical scale of $\sim$ 100 -- 800 pc down 
to masses of $\sim$ $10^{8}$ $M_{\odot}$ (see Table~\ref{tab:params}). 
Inside the bulge effective radius 
the surface brightness distribution for most of the sample galaxies is dominated by the
bulge contribution ($B/D$($r<r_{\rm e}$) $\gtrsim$ 1). 
Therefore, our luminosity-weighted spectroscopic measurements are probing the
bulge kinematics.

Considering a possible contamination from the underling disk dynamics,
to assure the reliability of the bulge velocity dispersion measurements
we create mock spectra of galaxies with bulge and disk components 
having physical properties that resemble our sample bulges.
We use the bulge half-light radius $r_{\rm e} = 0.35$ kpc, bulge S\'ersic index $n=1.2$, disk scale length $h=3$ kpc,
and bulge velocity dispersion $\sigma_{\rm e} = 58$ km s$^{-1}$.
We use several velocity fields for the disk component using the rotation curve profiles in \citet{Persic1996}.
To test some extreme possible scenarios,
we used values of disk velocities rising to more than 100 km s$^{-1}$ within the bulge half-light radius.
The value of the bulge velocity dispersion is reliably measured in all mock spectra,
with a small bias up to $\sim$5 km s$^{-1}$.

The dynamical status of the sample bulges is compared with 
a complementary sample of virialized stellar systems
over seven orders of magnitude in mass, comparing their physical properties using kinematics values
measured integrating the light within their half-light radius.
This represents a dissimilarity compared to the majority of previous works, where the central velocity dispersion of the galaxy is used 
instead of the value integrated within the half-light radius due to either the lack of the photometric information
or the less demanding spectroscopic data needed to retrieve the central value. 
The photometric decomposition of the galaxy light is indeed necessary to
accurately characterize various substructures (i.e., bulges).

Firstly, we complement the sample with the seven small bulges presented in \citet{Costantin2017}.
Then, we measure kinematics within the half-light radius and stellar masses of massive systems like large bulges \citep{MendezAbreu2017} 
and elliptical galaxies \citep{FalconBarroso2017, Iodice2019} using integral-field data from  
Calar Alto Legacy Integral Field Area Survey \citep[CALIFA;][]{Sanchez2016} and FORNAX3D \citep{Sarzi2018} samples, respectively.
Finally, we use literature values of kinematics and stellar masses of globular clusters \citep[GCs;][]{Taylor2015}, 
nuclear star clusters \citep[NSCs;][]{Kacharov2018},
compact early-type galaxies \citep[cETGs;][]{Norris2014, Guerou2015}, 
and ultra compact dwarf galaxies \citep[UCDs;][]{Mieske2008}. 
When only dynamical masses were available, we used them as upper limit for the stellar mass.

\section{Mass--velocity dispersion relation} \label{sec:results}

\begin{figure*}[t!]
\centering
\includegraphics[scale=0.48, trim=2.2cm 0.5cm 2.5cm 3cm, clip=true]{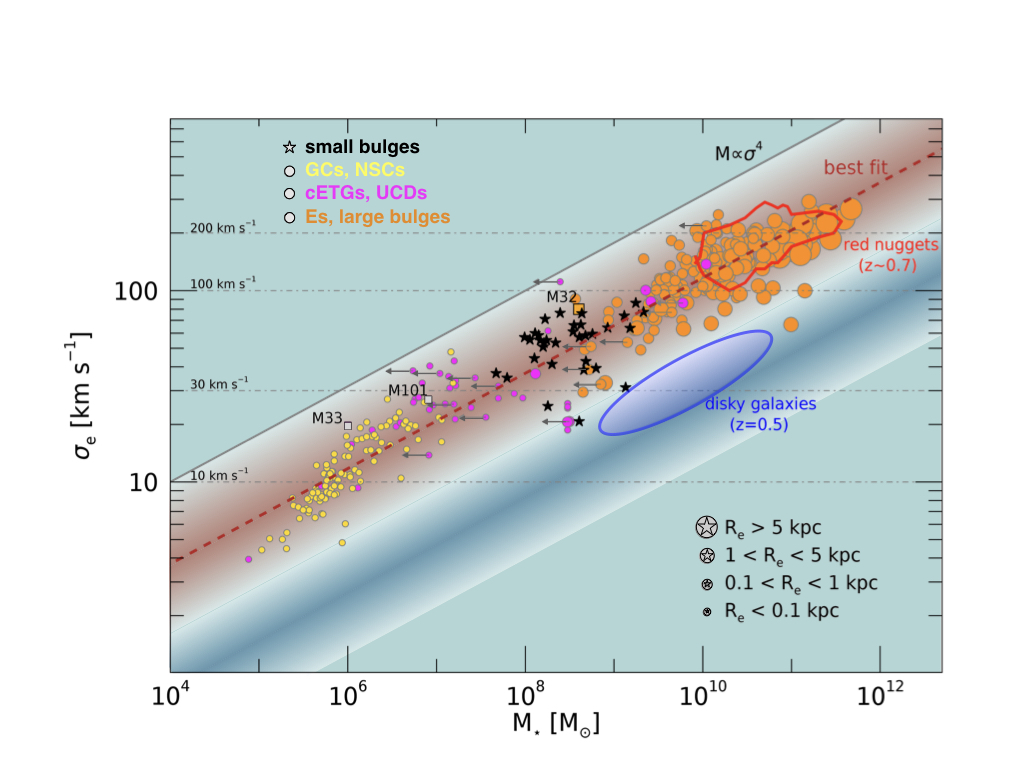}
\caption{Virial stellar mass versus velocity dispersion
(``mass Faber--Jackson'' relation) relation for small (globular clusters and nuclear stellar clusters;
yellow circles), compact (compact early type galaxies and ultra compact dwarfs; purple circles), and large
(ellipticals and large bulges; orange circles) systems.
Observations of small bulges are shown as black
stars. The best-fit relation (dark red dashed line) and the virial $M \propto \sigma^4$ relation (gray solid line) are shown. The
symbol size represents the effective radius of each system. The red shadow represents virialized stellar systems, where
the red contour marks the region of massive compact systems population (i.e., red nuggets) measured at redshift
$z \sim 0.7$. The blue shadow stands for the off-set sequence expected for disk-like stellar systems, where an example
at redshift $z = 0.5$ from Illustris TNG50 simulations is shown as blue contour. As a reference, 
the core of the spiral galaxies M33 and M101 and the compact elliptical galaxy M32 are superimposed to the main relation
(squares). Gray arrows stand for systems where dynamical masses are computed, which represent an upper limit for
the stellar mass.}
\label{fig:final_plot}
\end{figure*}

The photometric and kinematic information allows us to characterize the dynamical status of the different
stellar systems, including our small bulges, by means of their positions in fundamental scaling relations.
In Fig.~\ref{fig:final_plot} we show the stellar mass versus velocity dispersion (``mass Faber--Jackson'' relation), 
which is a direct consequence of the Virial Theorem and represents a projection of the Fundamental Plane.
This plane represents the \emph{cooling diagram} in theories of galaxy formation, 
relating the the virial temperature and the stellar density.
From a theoretical perspective, the amount of energy dissipation during the system formation 
and the timing at which it occurs is directly related to its position in this diagram. 
From the Virial Theorem, we can express the stellar luminosity as a function of the total mass as
\begin{equation}
L \propto \frac{r_{\rm e} \sigma^2}{(M/L)_{\rm vir}} \, ,
\end{equation}
where $r_{\rm e}$ and $\sigma$ are the half-light radius and the velocity dispersion of the galaxy, respectively.
Using the average surface brightness within the half-light radius ($\langle I_{\rm e} \rangle$),
the luminosity is $L\propto \langle I_{\rm e} \rangle r_{\rm e}^2$.
Considering that the luminosity is a proxy for the stellar mass, it is 
\begin{equation}
M_{\star} \propto \frac{\sigma^4}{\langle I_{\rm e} \rangle (M/L)^2_{\rm vir}} \, ,
\end{equation}
which corresponds to the ``mass Faber--Jackson'' relation if $\langle I_{\rm e} \rangle$ and $(M/L)_{\rm vir}$ are constant. 

We find that a tight relation (scatter of 0.15 dex) unifies large and more massive stellar systems
like elliptical galaxies ($r_{\rm e}$ $\gtrsim$ 5 kpc, $M_{\star} > 10^{11}$ $M_{\odot}$) 
and large bulges ($r_{\rm e}$ $\gtrsim$ 1 kpc, $M_{\star} = 10^{10}$ $M_{\odot}$, mostly residing in ETGs), 
and small and less massive stellar systems 
like NSCs ($r_{\rm e}$ $\sim$ 10 pc, $M_{\star} \sim$ $10^7$ $M_{\odot}$) 
and GCs ($r_{\rm e}$ $\sim$ 10 pc, $M_{\star} \sim$ $10^6$ $M_{\odot}$).
For some systems their dynamical masses are used as upper limits for the stellar mass.
Figure~\ref{fig:final_plot} shows that small bulges in late-type galaxies 
connect the previous stellar virialized systems over seven magnitudes in mass.
They bridge the gap between less and more massive virialized stellar systems,
sharing the same position in the plane as some compact systems like cETGs, UCDs, and massive NSCs.
As a reference, we also overplot the core of the spiral galaxies M33 and M101 \citep{Kormendy2010}
and the compact elliptical galaxy M32 \citep{Chilingarian2007}, which are shown to follow the relation. 
M32 is the closest (low-mass) elliptical galaxy,
while M33 and M101 represent two examples of nearby pure-disk galaxies (with a very small core).

Once we properly take into account the dynamical status of these systems 
by deriving their properties in a consistent way within the half-light radius, 
there is a remarkable agreement between the theoretical prediction from the Virial Theorem
and our best-fit estimation:
\begin{equation}
M_{\star} \propto \sigma_{\rm e}^{4.01 \pm 0.06}  \, .
\end{equation}
\citet{Cappellari2013} calculated $M_{\rm vir} \propto \sigma^{4.7 \pm 0.6}$ for galaxies with $\sigma \gg 140$ km s$^{-1}$ and
$M_{\rm vir} \propto \sigma^{2.3 \pm 0.1}$ for galaxies with $\sigma \ll 140$ km s$^{-1}$ in ATLAS$^{3D}$ sample.
\citet{Bezanson2015} derived $M_{\star} \propto \sigma^{2.44}$ using 
SDSS galaxies ($M_{\star} \sim 10^{10-11.5}$ $M_{\odot}$), 
with half-light radius velocity dispersion measurements derived applying aperture correction from the $3''$ SDSS fibers.
At higher redshift, analyzing a sample of DEIMOS galaxies ($M_{\star} \sim 10^{10-11.5}$ $M_{\odot}$) at $z\sim0.7$, 
\citet{Bezanson2015} obtained $M_{\star} \propto \sigma^{2.94}$.
All these results agree with \citet{Balcells2007} ($L \propto \sigma^{2.9 \pm 0.5}$) and 
\citet{Costantin2017}  ($L \propto \sigma^{2.64 \pm 0.01}$).
We suggest these differences from the theoretical prediction are due to either
the small mass range explored or the not homogeneous/reliable comparison sample.
By exploring a wider mass regime as we have done for the first time in this work,
which also requires accurate kinematic measurements, we replicate
the Virial Theorem prediction describing the physical resemblance
between all these virialized systems in the ``mass Faber--Jackson'' relation.

Finally, we compare the kinematics of the virialized systems 
with the one of disk galaxies at $z = 0.5$, as predicted from the state-of-the-art 
hydrodynamical simulations \citep[Illustris TNG50;][]{Pillepich2019}. 
Disk galaxies are offset from the virial relation, ruling out 
the possibility that the dynamical status of our small bulges resembles the one of disk-like systems.

\section{Discussion and Conclusions\label{sec:discussion}}

The relation between fundamental galactic dynamical and structural properties, which manifests in global scaling laws,
carries important clues about the process of galaxy formation in a cosmological context.
Interpreting the phenomenology of these relations can indeed have 
substantial consequences on our understanding of the interplay 
between baryons and dark matter, including its varying nature in the $\Lambda$CDM paradigm.

We find that stellar systems over seven orders of magnitude in mass follow
the same fundamental ``mass Faber--Jackson'' relation, which directly follows from the Virial Theorem prediction.
This result is achieved by homogeneously comparing
systems whose physical properties are measured within their half-light radius
using reliable spectroscopic information.
While the observed trend is not surprising for elliptical galaxies and massive bulges,
we reveal that small bulges are not offset from this relation and follow the $M_{\star} \propto \sigma^4$ prediction from the Virial Theorem.
This leads to our new interpretation that small bulges in the center of today's extremely late-type
galaxies maintain memory of an ancient formation, as virialized systems do.
Moreover, small bulges share the same position in the ``mass Faber--Jackson'' plane as cETGs, UCDs, and massive NSCs.
This further support our proposed scenario, because (at least) the most massive cETGs, UCDs, and GCs (e.g., $\omega$Cen in the Milky Way)
have been proposed to be linked to the remnants of tidally disrupted nucleated galaxies \citep{Drinkwater2003},
which represent the first generation of cluster satellites and are the result of an early 
and rapid formation \citep{SanchezJanssen2019}.
This early formation of the central region of late-type galaxies alleviate the tension between
SMBH and bulge coevolution at high redshift. Thus, small bulges should not represent an exception of
the fundamental scaling relations that regulate SMBH and bulge interplay, 
even though the progressively lower binding energies of low-mass galaxies intuitively 
allow for a different regulation of SMBH and bulge synergic growth.

Recently, using N-body simulations Guo et al.~(2019) have shown that the destruction of short bars
due to the presence of SMBH of $\sim0.1$\% the stellar mass in galaxies 
of $M_{\star} \sim 4\times10^{10}$ $M_{\odot}$, could end up mimicking the photometric and kinematic properties
of virialized bulges.
However, the fraction of galaxies experiencing the different phases required by this scenario
(early inner disk formation, short bar formation, and destruction) 
is difficult to estimate with the available observations and simulations.

On the other hand, the scenario proposed by \citet{Hopkins2009} 
postulates that the massive high-redshift quiescent compact galaxies \citep[red nuggets;][]{Damjanov2009}
end up in the center of present-day massive elliptical galaxies.
This traditional evolutionary picture leads to extend the paradigm to disk galaxy bulges \citep{Graham2013, delaRosa2016}. 
In the $\Lambda$CDM context, \citet{Zolotov2015} used cosmological simulations to show that after compaction
extended star-forming disks can develop around red nuggets.
It is worth noting that the velocity dispersion of the central component in galaxies is
relatively unaffected by the loss of an outer stellar or dark matter halo \citep{Chilingarian2009}, 
and seems also not to be perturbed by the growth of the extended disk component \citep{DeBattista2013}.
The results presented in this work support the idea that also late-type galaxies might 
host compact stellar systems in their central region,
providing a general and broader picture of galaxy evolution at all masses in an inside-out scenario.
Therefore, to further investigate a possible connection between galaxies in the Local universe
and their progenitors, we overplot in Fig.~1 a sample representing the red nugget population observed by \citet{Bezanson2015} 
at $z\sim0.7$, showing that they occupy the massive end of the ``mass Faber--Jackson'' relation.
In this view, we speculate that not only the center of ETGs should hide the remnants of the compact 
objects population observed at high redshift, but also
the core of late-type galaxies 
could represent the low-mass tail of high-redshift formed red nuggets.

\acknowledgments
{\small We would like to thank the anonymous referee for improving the content of the manuscript.
LC wishes to thank CCG for inspiration and AC for the useful discussion while this manuscript was written.
LC acknowledges financial support from Comunidad de Madrid under Atracci\'on
de Talento grant 2018-T2/TIC-11612 and the Spanish Ministerio de Ciencia, Innovación y Universidades through grant PGC2018-093499-B-I00.
JMA and AdLC acknowledge support from the Spanish Ministerio de Economia y Competitividad (MINECO) 
grants AYA2017-83204-P and AYA2016-77237-C3-1-P, respectively.
EMC is supported by MIUR grant PRIN 2017 20173ML3WW\_001. 
IP is supported by Leibniz-Institut fur Astrophysik Potsdam.
VC and IP acknowledge support from the Fondazione Ing.~Aldo Gini. 
EMC, VC, and MR are funded by Padua University grants DOR1715817/17, DOR1885254/18, and DOR1935272/19.
}

\bibliography{Costantin}{}
\bibliographystyle{aasjournal}

\end{document}